\documentstyle[12pt]{article}
\newcommand{\text}[1]{\mbox{#1}}

\begin{document}
\vspace*{-60pt}
\begin{flushright}
DPNU-97-52 \\
October 1997
\end{flushright}
\bigskip\bigskip
\begin{center}
{\LARGE {\bf Quantum Mechanics on $S^{n}$ \\
    and Meron Solution}}
\end{center}

\vspace{30pt}

\begin{center}
Hitoshi IKEMORI\footnote{%
ikemori@biwako.shiga-u.ac.jp} \\[0pt]
{\it Faculty of Economics, Shiga University,} {\it Hikone, Shiga 522, Japan}

\vspace{10pt}

Shinsaku KITAKADO\footnote{%
kitakado@eken.phys.nagoya-u.ac.jp} \\[0pt]
{\it Depertment of Physics, Nagoya University,} {\it Nagoya 464-01, Japan}

\vspace{10pt}

Hajime NAKATANI \\[0pt]
{\it Chubu Polytechnic Center, Komaki, Aichi 485, Japan}

\vspace{10pt}

Hideharu OTSU\footnote{%
otsu@vega.aichi-u.ac.jp} \\[0pt]
{\it Faculty of General Education, Aichi University, }

{\it Toyohashi, Aichi 441, Japan}

\vspace{10pt}

Toshiro SATO\footnote{%
tsato@matsusaka-u.ac.jp} \\[0pt]
{\it Faculty of Political Science and Economics,}\\[0pt]
{\it Matsusaka University, Matsusaka, Mie 515, Japan}
\end{center}

\vspace*{30pt}

\begin{center}
{\Large Abstract}
\end{center}

A particle in quantum mechanics on manifolds couples to the 
induced topological gauge field that characterises the possible 
inequivalent quantizations.
 For instance, the gauge potential induced on $S^2$ is that 
of a magnetic monopole located at the center of $S^2$.
 We find that the gauge potential induced on 
$S^3$ ($S^{2n+1}$) is that of a meron (generalized meron) 
also sitting at the center of $S^3$ ($S^{2n+1}$).

\newpage

In recent years, it has been noted in various situations that the
topological gauge structure may be induced upon quantization on
topologically nontrivial manifolds.\cite{Mackey}\cite{Ohnuki-Kitakado92-93}%
\cite{McMullan-Tsutsui} This gauge strucure characterises the inequivalent
quantizations that are possible on these manifolds. Particularly clear
results are obtained for quantum mechanics on $S^{n}$ with radius $r$
embedded in $R^{n+1}$.\cite{Ohnuki-Kitakado92-93}

Let $x_{\alpha }$ $\left( \alpha =1,2,\ldots ,n+1\right) $ be a homogenious
coordinate of $R^{n+1}$ and define the $S^{n}$ by the constraint

\begin{equation}
\sum_{\alpha =1}^{n+1}x_{\alpha }^{2}-r^{2}=0.
\end{equation}
We shall represent the observables by the use of the covariant form in $%
R^{n+1}$ that should be projected on the $S^{n}.$

Then the fundamental algebra of observables on the $S^{n}$ is given by \cite
{Ohnuki-Kitakado92-93} 
\begin{eqnarray}
\left[ x_{\alpha },x_{\beta }\right] &=&0, \\
\left[ x_{\lambda },G_{\alpha \beta }\right] &=&i\left( x_{\alpha }\delta
_{\lambda \beta }-x_{\beta }\delta _{\lambda \alpha }\right) ,  \nonumber \\
\left[ G_{\alpha \beta },G_{\lambda \mu }\right] &=&i\left( \delta _{\alpha
\lambda }G_{\beta \mu }-\delta _{\alpha \mu }G_{\beta \lambda }+\delta
_{\beta \mu }G_{\alpha \lambda }-\delta _{\beta \lambda }G_{\alpha \mu
}\right) ,  \nonumber
\end{eqnarray}
and the induced gauge potential that appears in this representation has the
form for $n\geq 2$%
\begin{equation}
\left\{ 
\begin{array}{c}
A_{j}\left( x\right) =\displaystyle\frac{1}{r(r+x_{n+1})}%
\sum_{k=1}^{n}S_{jk}x_{k},\;\left( j,k=1,2,...n\right) \\ 
A_{n+1}\left( x\right) =0
\end{array}
\right. ,
\end{equation}
where $S_{jk}$'s are Hermitian matrices of $SO\left( n\right) $. These can
be written in $SO(n+1)$ covariant form

\begin{equation}
A_{\alpha }=\frac{1}{r^{2}}\sum_{\beta =1}^{n+1}S_{\alpha \beta }x_{\beta
}\;,
\end{equation}
with the constraint on the wave function

\begin{equation}
\frac{1}{r}\sum_{\alpha =1}^{n+1}x_{\alpha }\gamma _{\alpha }\psi =\pm \psi 
\text{ }.
\end{equation}
\smallskip Where $\gamma _{\alpha }$'s are the Hermitian matrices satisfying
the Clifford algebra of the order $n+1$ and $S_{\alpha \beta }$'s are
Hermitian matrices that satisy the Lie algebra of $SO(n+1)$ and can be
written as

\begin{equation}
S_{\alpha \beta }=\frac{1}{4i}\left[ \gamma _{\alpha },\gamma _{\beta
}\right] .
\end{equation}

The potentials induced in the cases $n=1$ and $n=2$ can be visualised as
those resulting from a magnetic flux and a magnetic monopole respectively
located at the center of the embedding space. The even $n$ cases for $n\geq
4 $ were interpreted as the (generalized) instantons,\cite{McMullan-Tsutsui} 
\cite{Fujii-Kitakado-Ohnuki}which are the habitants on $S^{n}$. On the other
hand, there are no topological interpretations for the odd $n$ $\left( n\geq
3\right) $ cases, the first question in this respect being `` What is
sitting at the center of $S^{3},S^{5},S^{7},\dots $ ?''. In this note we
shall see that the odd $n$ $\left( n\geq 3\right) $ cases can similarly be
considered to be the potentials generated by the (generalized) merons
sitting at the center of the embedding space $R^{n+1}$. Although the meron
solutions have singularities, they exist in our picture outside of the
physical $S^{n}$, i.e. at the center of $R^{n+1}$ wherein the former is
embedded. Thus we can expect the meron effect on the surface. The authors of
ref.\cite{Hirayama-Zhang-Hamada} have explicitly obtained the generators
satisfying the fundamental algebra in terms of the induced gauge potential
and have discovered among others the zero size instanton for $n=3$ at the
center of $S^{3}$. On the other hand, in ref.\cite{McMullan-Tsutsui}\cite
{Fujii-Kitakado-Ohnuki} BPST instantons appear on $S^{2n}$, $n=2,3,...,$
embedded in $R^{2n+1}$ .

In case of $S^{3}$, the authors of ref.\cite{Ohnuki-Kitakado92-93} have
obtained an induced gauge field, without argument on topological
interpretations. Their gauge field can be written modulo gauge
transformations (in $\gamma _{5}$-diagonal represenation) as follows,

\begin{equation}
A_{\mu }=\frac{1}{r^{2}}S_{\mu \nu }x_{\nu }=\frac{1}{r^{2}}\left( 
\begin{array}{ll}
\sigma _{\mu \nu } & 0 \\ 
0 & \overline{\sigma }_{\mu \nu }
\end{array}
\right) x_{\nu }\text{ },
\end{equation}
where $S_{\mu \nu }$'s $\left( \mu ,\nu =1,2,3,4\right) $ are the generators
of the $SO\left( 4\right) $ and 
\begin{equation}
\sigma _{ij}=\overline{\sigma }_{ij}=\frac{1}{2}\epsilon _{ijk}\tau
_{k},\;\sigma _{i4}=-\overline{\sigma }_{i4}=\frac{1}{2}\tau _{i}\;\left(
\tau _{i}:\text{Pauli matrices}\right) .
\end{equation}
Introducing the four component spinor

\begin{equation}
\psi =\left( 
\begin{array}{c}
\psi _{R} \\ 
\psi _{L}
\end{array}
\right) ,
\end{equation}
where 
\[
\psi _{\matrix{\scriptstyle R \cr  \raise1.4ex\hbox{$\scriptstyle L$} }}
=\frac{1\pm \gamma _{5}}{2}\psi , 
\]
the constraint (5) reduces to

\begin{equation}
\psi _{L}=\pm g^{-1}(x)\psi _{R},
\end{equation}
where 
\begin{equation}
g(x)=\frac{\left( x_{4}-i\vec{x}\cdot \vec{\tau}\right) }{r}\;\;\left( r=%
\sqrt{x_{1}^{2}+x_{2}^{2}+x_{3}^{2}+x_{4}^{2}}\right) .
\end{equation}
Thus our wave functions are essentialy two component spinors and we can
rewrite (7) as

\begin{equation}
A_{\mu }=\frac{1}{r^{2}}\sigma _{\mu \nu }x_{\nu },
\end{equation}
or 
\begin{equation}
A_{\mu }=\frac{1}{r^{2}}\bar{\sigma}_{\mu \nu }x_{\nu },
\end{equation}
which can be considered as $SU(2)$ Yang-Mills (Y-M) gauge fields. Note that
there is a case of the trivial gauge field modulo gauge transformations 
\begin{equation}
A_{\mu }=0,
\end{equation}

We shall clarify that we can regard these configurations as topologically
nontrivial objects in the embedding space $R^{4}.$ These gauge fields can be
considered as $SU\left( 2\right) $ gauge fields in $R^{4}$ space and they
satisfy the Y-M equation in the embedding space.

Recall that the field equation 
\begin{equation}
D_{\mu }f_{\mu \nu }=\partial _{\mu }f_{\mu \nu }-i\left[ a_{\mu },f_{\mu
\nu }\right] =0,
\end{equation}
for the 4-dimensional Euclidean $SU\left( 2\right) $ Y-M theory has
instanton solutions,\cite{Actor} the instanton number of which is 
\begin{equation}
Q=\frac{1}{16\pi ^{2}}\int {\rm tr}\left( f_{\mu \nu }\tilde{f}_{\mu \nu
}\right) d^{4}x=0,\pm 1,\pm 2,\ldots \;,
\end{equation}
due to $\Pi _{3}\left( SU(2)\right) =Z$. The $Q=0$ case corresponds to the
trivial solution with 
\begin{equation}
a_{\mu }=0\;,
\end{equation}
the single instanton solution $(Q=1)$ can be written as 
\begin{equation}
a_{\mu }=\frac{2}{r^{2}+\lambda ^{2}}\sigma _{\mu \nu }x_{\nu },
\end{equation}
and we also have the multi-instanton solutions with $Q=m$.

In addition to these, there is a solution of Y-M equation called ``meron'' 
\begin{equation}
a_{\mu }=\frac{1}{r^{2}}\sigma _{\mu \nu }x_{\nu },
\end{equation}
which has half an instanton number.\cite{Actor}

Let us now consider each cases of induced gauge fields separately. First we
consider the case of (14) $A_{\mu }=0.$ This is a trivial configuration,
however, we note that we are concerned with the $SU(2)$ gauge fields and in
addition to the small gauge transformations there are large gauge
transformations, that can change the instanton number. The large gauge
tansformation with the winding number 1 is 
\[
g(x)=\frac{\left( x_{4}-i\vec{x}\cdot \vec{\tau}\right) }{r}\;, 
\]
while those with the winding number $m$ are 
\[
g^{m}\;(x)\left( m=\pm 1,\pm 2,\ldots \right) . 
\]
Thus we can have the induced gauge fields due to large gauge transformation 
\begin{equation}
A_{\mu }=ig^{-1}(x)\partial _{\mu }g(x),
\end{equation}
which could be considered as the $\lambda \rightarrow 0$ limit of an
instanton or two merons sitting at $r=0$. This is the configuration found in
ref.\cite{Hirayama-Zhang-Hamada} and the instanton number inside the $S^{3}$
is 
\begin{equation}
Q_{\text{inside}}=1.
\end{equation}

Next we consider the case of (12)\footnote{%
Note that (13) can be considered as the case of $m=-1$ in (27).} 
\[
A_{\mu }=\frac{1}{r^{2}}\sigma _{\mu \nu }x_{\nu }. 
\]
In fact this is the meron solution inferred in (19), the instanton density
of which is $\frac{1}{2}\delta \left( r\right) $ . This field is specified
by the instanton number 
\begin{equation}
Q_{\text{inside}}=\frac{1}{2}\text{ }.
\end{equation}

Thus in general 
\begin{equation}
A_{\mu }=0+\left( \text{large gauge trans.}\right) ,
\end{equation}
can be expressed as 
\begin{equation}
A_{\mu }=ig^{-m}(x)\partial _{\mu }g^{m}(x),
\end{equation}
and 
\begin{equation}
Q_{\text{inside}}=m\text{ }.
\end{equation}
On the other hand, 
\begin{equation}
A_{\mu }=\frac{1}{r^{2}}\sigma _{\mu \nu }x_{\nu }+\left( \text{large gauge
trans.}\right) ,
\end{equation}
reduces to 
\begin{equation}
A_{\mu }=g^{-m}(x)\frac{1}{r^{2}}\sigma _{\mu \nu }x_{\nu
}g^{m}(x)+ig^{-m}(x)\partial _{\mu }g^{m}(x),
\end{equation}
and 
\begin{equation}
Q_{\text{inside}}=\frac{1}{2}+m.
\end{equation}

This argument can of course be generalized to $S^{2n-1}\;\left( n\geq
2\right) .$ We again have two cases with

\begin{equation}
A_{\mu }=iG^{-m}\partial _{\mu }G^{m},
\end{equation}
\begin{equation}
Q_{\text{inside}}=m,
\end{equation}
and 
\begin{equation}
A_{\mu }=G^{-m}\frac{1}{r^{2}}\Sigma _{\mu \nu }x_{\nu
}G^{m}+iG^{-m}\partial _{\mu }G^{m},
\end{equation}
\begin{equation}
Q_{\text{inside}}=\frac{1}{2}+m,
\end{equation}
where, $\Sigma _{\mu \nu }$ is the generator of $SO\left( 2n\right) $, which
can be expressed in terms of $\Gamma _{\mu }$ ($\mu =1,2,...,2n$) satisfying
the Clifforg algebra of order $2n$ as 
\begin{equation}
\Sigma _{\mu \nu }=\frac{1}{4i}\left( \frac{1+\Gamma _{2n+1}}{2}\right)
\left[ \Gamma _{\mu },\Gamma _{\nu }\right] .
\end{equation}
$G$ is large gauge transformation that follows from nontriviality of $\Pi
_{2n-1}\left( SO(2n)\right) .$ $m=0$ case of (31) is nothing but the
(generalized) meron solution \cite{Brien-Tchrakian} of YM theory in $R^{2n}$.

To summarize, the induced gauge fields for quantum mechanics on $S^{2n-1}$ $%
(n\geq 2)$ can be considered to be generated by merons (anti-merons) sitting
at the center ($r=0$) of $R^{2n}$ where our $S^{2n-1}$ is embedded. The
gauge fields are specified by the instanton number $Q_{\text{inside}}=\frac{m%
}{2}\;$($m$:integer). For even $m$ , $F_{\mu \nu }=0$ on $S^{2n-1}$, but for
odd $m$, $F_{\mu \nu }\neq 0$ even on $S^{2n-1}$.

To see the physical effect to a particle on $S^{3}$ of a meron at the
center, let us consider a quantum mechanical phase factor 
\begin{equation}
P\exp \left[ i\oint A_{\mu }dx^{\mu }\right] ,
\end{equation}
for the closed loop. To be specific we consider the case $x_{4}=0$, i.e. the
closed loop inside $S^{2}$ $(x_{1}^{2}+x_{2}^{2}+x_{3}^{2}=r^{2})$. In the
region $x_{4}=0$ the meron solution

\begin{equation}
A_{\mu }=\frac{1}{r^{2}}\sigma _{\mu \nu }x_{\nu },
\end{equation}
reduces to the Wu-Yang monopole 
\begin{equation}
A_{i}=\frac{1}{2r^{2}}\epsilon _{ijk}x_{j}\tau _{k},
\end{equation}
with the magnetic charge $g=1/2e$. Thus the phase factor is fixed by the
magnetic flux going through the loop on $S^{2}$ that comes from the
monopole. A particle on $S^{3}$, moving in the region of $x_{4}=0$, is under
the same effect as a particle on $S^{2}$ with a monopole of the magnetic
charge $g=1/2e$. On the other hand, for quantum mechanics on $S^{3}$ with an
instanton(2-merons) at the center, no physical effect is expected on $S^{3}$
at least for $x_{4}=0$, since no magnetic flux pierces through $S^{2}$.

Recent investigations have revealed a close connection between the quantum
mechanically induced gauge potentials on manifolds on one hand and the so
called topological terms appearing in field theories on the other.\cite{IKOS}%
\cite{Kobayashi-Tsutsui-Tanimura}\cite{Miyazaki-Tsutsui} Actually, the Hopf
term in the $2+1$ dimensional $O(3)$ nonlinear sigma model in $R^{2}$-space
has been shown \cite{IKOS}\cite{Kobayashi-Tsutsui-Tanimura} to be an effect
of quantum mechanics on $S^{1}$, while the Wess-Zumino term in the chiral
model defined in $1+1$ dimensions can be understood \cite{Miyazaki-Tsutsui}
as the Dirac monopole potential that is induced on $S^{2}$. It is
interesting to note that $O(3)$ nonlinear sigma model on $S^{2}$ space \cite
{IKOS} were shown to be described by quantum mechanics on $S^{3}$ where the
merons studied in this note should play a role.

\vspace{30pt}

{\Large Acknowledgments}{\large \ }{We would thank K. Hasebe fou useful
discussions. This work has been supported in part by the Grant-in-Aid for
Scientific Research No. 09640344.}

\end{document}